\documentclass[11pt,a4paper]{article}
\usepackage{jcappub}
\usepackage{amssymb}
\usepackage{graphicx}
\usepackage{amsmath}
\usepackage{amsfonts}
\usepackage{hyperref}

\title{Modified Einstein's gravity as a possible missing link between sub- and super-Chandrasekhar type~Ia supernovae}

\author[a]{Upasana Das}
\author[a]{and Banibrata Mukhopadhyay \footnote{Corresponding Author.}}

\affiliation[a]{Department of Physics, Indian Institute of Science, 
Bangalore 560012, India}

\emailAdd{upasana@physics.iisc.ernet.in}
\emailAdd{bm@physics.iisc.ernet.in}

\abstract{We explore the effect of modification to Einstein's gravity in white dwarfs
for the first time in the literature, to the best of our knowledge.
This leads to significantly sub- and super-Chandrasekhar limiting masses of white dwarfs,
determined by a single model parameter. 
On the other hand, type~Ia supernovae (SNeIa), a key to unravel the evolutionary history 
of the universe, are believed to be triggered in white dwarfs having mass close to the Chandrasekhar 
limit. However, observations of several peculiar, under- and over-luminous 
SNeIa argue for exploding masses widely different from this limit. 
We argue that explosions of the modified gravity induced sub- and super-Chandrasekhar limiting mass white dwarfs 
result in under- and over-luminous SNeIa respectively, thus unifying these
two apparently disjoint sub-classes and, hence, serving as a missing link.
Our discovery raises two fundamental questions. Is the Chandrasekhar limit unique? 
Is Einstein's gravity the ultimate theory for understanding astronomical phenomena? 
Both the answers appear to be no!}


\arxivnumber{1411.1515}

\begin{document}
\maketitle


\section{Introduction} 

Modification to Einstein's theory of gravity, although has been applied to neutron stars,
has never been explored for white dwarfs. Perhaps the reason for not doing so, until the present work,
is the larger size and lower density of white dwarfs, which apparently argue for their much weaker gravitational
field compared to neutron stars. In this work, we aim at exploring the effect of modification to Einstein's gravity 
in white dwarfs. This shows that modified gravity effect is quite non-negligible in high density white dwarfs
and could be significant depending on the value of a model parameter.
We furthermore argue that our result may have far reaching astrophysical implications.

It has been understood that our universe exhibits accelerated expansion, 
which has been firmly established by
the observations of extremely luminous stellar explosions, 
known as type~Ia supernovae (SNeIa). SNIa is 
one of the most widely studied astronomical events. 
These SNeIa are believed to result from the
violent thermonuclear explosion of a carbon-oxygen white dwarf,
when its mass approaches the famous Chandrasekhar limit of
$1.44M_\odot$, where $M_\odot$ is the solar mass. The characteristic nature of the
variation of luminosity with time of SNeIa is believed to be powered by the decay of
$^{56}\!$Ni to $^{56}\!$Co and, finally, to $^{56}\!$Fe. This feature, along
with the consistent mass of the exploding white dwarf, allows
SNeIa to be used as a `standard' for measuring far
away distances (standard candle) and, hence, in understanding
the expansion history of the universe \cite{perl99}.

However, the discovery of several peculiar SNeIa
provokes us to rethink the commonly accepted scenario.
Some of these SNeIa are highly over-luminous, e.g.  SN 2003fg, SN 2006gz, SN 2007if, SN 2009dc \cite{howel,scalzo},
and some others are highly under-luminous, e.g. SN 1991bg, SN 1997cn, SN 1998de, SN 1999by, SN 2005bl
\cite{1991bg,mazzali97,turato97,modjaz,garnavich,taub2008,gonz} (see also \cite{tutu}). 
The luminosity of the former group of SNeIa (super-SNeIa) implies a huge Ni-mass 
(often itself super-Chandrasekhar), invoking highly super-Chandrasekhar 
white dwarfs, having mass $2.1-2.8M_\odot$, as their most plausible progenitors \cite{howel,scalzo,taub,yam,hicken,silverman}. 
On the other hand, the latter group (sub-SNeIa) produces as low
as $\sim 0.1M_\odot$ of Ni \cite{stritz}. 
Attempted models to explain sub-SNeIa, often based on numerical simulations, include 
explosion due to the merging of two sub-Chandrasekhar white dwarfs \cite{pakmor-nature},
explosion of a single sub-Chandrasekhar white dwarf triggered externally due to accretion of a helium layer 
(sub-Chandrasekhar mass model) \cite{hillebrandt}.
However, they entail caveats, such as, the simulated light-curve in the merger scenario 
fades slower than that suggested by observations \cite{pakmor-nature}, along with other spectroscopic discrepancies 
in individual models \cite{mazzali2012}.
The models, in order to explain super-SNeIa progenitor mass, include rapidly (and differentially) 
rotating white dwarfs \cite{yoon}, binary evolution of accreting, differentially rotating white dwarfs \cite{hachisu}, 
highly magnetized white dwarfs \cite{dmprl,dmapjl}. However, they also 
harbor several doubts such as, existence of supermassive ($>1.7M_\odot$), stable, highly rotating white dwarfs \cite{chen}, 
stability of highly magnetized white dwarfs \cite{chamel}. Nevertheless, the issues related to highly magnetized
white dwarfs have been addressed, e.g., by considering varying magnetic fields within them \cite{mpla14,jcap}.

Even if we keep aside all the aforementioned caveats, a major concern arises that such a large array
of models is required to explain apparently the same phenomena, i.e., triggering of thermonuclear explosions
in white dwarfs. It is unlikely that nature would seek mutually antagonistic scenarios to exhibit 
sub- and super-SNeIa, which are sub-classes of the same SNeIa. 
This is where the current work steps in. Our work attempts to unify 
the phenomenologically disjoint sub-classes of SNeIa by effectively a single underlying theory, which, hence,
serves as a missing link. This is achieved by invoking 
a modification to Einstein's theory of gravity or general relativity in white dwarfs. 

The validity of general relativity has been tested mainly in the weak field regime,
for example, through laboratory experiments and solar system tests. The expanding universe,
the region close to a black hole and neutron stars are the
regimes of strong gravity. The question is, whether general relativity is the ultimate theory of
gravitation, or it requires modification in the strong gravity regime. It is important to note that
a modified theory of gravity, which explains observations that general relativity cannot
(as we will demonstrate here),
should also be able to reproduce observations in the regime
where general relativity is adequate (which we will establish below as well). 
Indeed, it was shown long back that such modified gravity theories reveal significant deviations to
the general relativistic solutions of neutron stars \cite{dam}.
As neutron stars are much more compact than white dwarfs, so far, modified gravity theories
have been applied only to them in order to test the validity of such theories in the strong
field regime. The current venture with white dwarfs is a first in the literature to the best of our
knowledge.

In the next section, we briefly recall the basic equations of general relativity and 
then move on to describe the modification of Einstein's theory which we invoke. 
Subsequently, we discuss the perturbative solution procedure employed in section \ref{solnp} and the final results 
obtained in section \ref{res}. 
Finally, we end with conclusions in section \ref{conc}.

\section{Basic equations and modified gravity model}

We mostly use geometrized units while deriving various 
equations, which is, $c=G=1$, unless otherwise mentioned, where $c$ is the speed of light and $G$ 
Newton's gravitation constant. We also use the metric signature $(-,+,+,+)$.

The standard method to arrive at an equation of motion in any field theory is 
applying a variational principle. In general relativistic field theory, one starts with the 
Einstein-Hilbert action in 4 dimensions given by \cite{wald}
\begin{eqnarray}
S &=& \int ({\cal L}_G +{\cal L}_M) \sqrt{-g}~ d^4 x \nonumber \\
 &=& \int \left[\frac{1}{16\pi} R + {\cal L}_M \right] \sqrt{-g}~ d^4 x , 
\label{action}
\end{eqnarray}
where $g$ is the determinant of the metric tensor $g_{\mu\nu}$ (which describes the nature
of the underlying curvature of spacetime), ${\cal L}_G$ the Lagrangian density 
of the gravitational field and ${\cal L}_M$ the Lagrangian density of the matter field. 
${\cal L}_G$ in general relativity is simply $R/16\pi$, where the Ricci scalar $R$ is defined by $R=g^{\mu\nu}R_{\mu\nu}$, 
when $R_{\mu\nu}$ is the Ricci tensor, which is defined as
\begin{equation}
R_{\mu\nu} = \partial_\lambda \Gamma^\lambda_{~~\mu\nu} - \partial_\nu \Gamma^\lambda_{~~\lambda\nu} +  \Gamma^\lambda_{~~\lambda\sigma} \Gamma^\sigma_{~~\mu\nu} -\Gamma^\lambda_{~~\sigma\nu} \Gamma^\sigma_{~~\lambda\mu} ,
\end{equation}
where $\Gamma^\lambda_{~~\mu\nu}=(1/2)g^{\lambda\sigma} (\partial_\mu g_{\nu\sigma} + \partial_\nu g_{\mu\sigma} - \partial_\sigma g_{\mu\nu})$, 
is known as the Christoffel symbol \cite{wald}.
If the above action, equation (\ref{action}), is varied with respect to 
$g_{\mu\nu}$ and then extremized such that $\delta S=0$, then one obtains the famous Einstein's field equations
\begin{equation}
G_{\mu\nu} = R_{\mu\nu} - \frac{1}{2}R~ g_{\mu\nu}=8\pi T_{\mu\nu} ,
\label{gr}
\end{equation}
where $G_{\mu\nu}$ is the Einstein tensor and 
$T_{\mu\nu}$ the energy-momentum tensor of the matter field. 

Now, in a modified gravity theory, the left hand side of equation (\ref{gr}) is modified 
leaving the right hand side unchanged. One such a very popular class of modified gravity 
theory is known as the $f(R)$ theory, in which ${\cal L}_G$ is replaced by $f(R)/16\pi$, where $f$ is an 
arbitrary function of $R$. The action for $f(R)$ gravity is thus \cite{livrel,nojiri}
\begin{equation}
S = \int \left[\frac{1}{16\pi} f(R) + {\cal L}_M \right] \sqrt{-g}~ d^4 x , 
\end{equation}
varying which with respect to $g_{\mu\nu}$, one arrives at the following modified field equation
\begin{equation}
F(R) G_{\mu\nu} + \frac{1}{2}g_{\mu\nu}(F(R) R - f(R))- (\nabla_\mu \nabla_\nu -g_{\mu\nu} \Box)F(R) = 8\pi T_{\mu\nu} ,
\label{modgr}
\end{equation}
where $F(R)\equiv \partial f(R)/\partial R$, the covariant derivative $\nabla_\mu$ 
acting on a vector $A_\nu$ is defined as $\nabla_\mu A_\nu = \partial_\mu A_\nu - \Gamma^\sigma_{~~\mu\nu} A_\sigma$ 
and the d'Alembertian operator 
$\Box \equiv \partial_\mu(\sqrt{-g} g^{\mu\nu}\partial_\nu)/\sqrt{-g}$.
Thus an $f(R)$ theory reduces to general relativity for $f(R)=R$ and, hence, $F(R)=1$.

For the present purpose, we choose the Starobinsky model \cite{starobi}
or the $R$-squared model of modified gravity defined as
\begin{equation}
f(R)=R+\alpha R^2 ,
\label{ar2}
\end{equation}
where $\alpha$ is a constant having the dimension of length-squared. 
Henceforth, in the rest of the article, by modified gravity effects we would mean the effects of Starobinsky model.
This model/theory can be conformally related to a scalar-tensor theory \cite{whitt}, and this connection has also been explored very recently in the context of quark stars \cite{quark}. These theories have been
furthermore tested against binary pulsar observations and solar system measurements. From such observations,
one obtains the bound on the linear coupling constant of the scalar-tensor theory, $\alpha_0$, to be few factor times $10^{-3}$ (e.g. \cite{dam,dam2,fre}).  
In the Starobinsky model, $|\alpha R|$ may be thought of as playing the role equivalent to $\alpha_0$, and we will show below that
the dimensionless model parameter $|\alpha R|$ in the present work has been indeed 
restricted to few factor times $10^{-3}$.
Moreover, the strictest astrophysical upper bound on the value of 
$\alpha$ itself has been set by the Gravity Probe B experiment, 
namely $|\alpha|\lesssim 5\times 10^{15}$ ${\rm cm^2}$ \cite{naf}. Hence, the value of $\alpha$ may not be chosen arbitrarily. 
In this context, we furthermore mention that bounds on $\alpha$ have been obtained from other systems as well. 
For example, the E\"{o}t-Wash laboratory experiment sets the 
limit as $\alpha \lesssim 10^{-6}$ ${\rm cm^2}$, while the precession of the pulsar B in the double-pulsar binary 
system PSR J0737-3039 gives $\alpha \lesssim 2.3\times 10^{19}$ ${\rm cm^2}$ \cite{naf}. This apparent 
huge difference in the bounds may be explained invoking an underlying 
``chameleon" effect \cite{cham2}. This in turn would cause $\alpha$ to vary depending on the characteristic length scale or density of the system under consideration.

However, similar effects, as of Starobinsky $f(R)$-model, could also be obtained in other modified gravity theories, e.g. Born-Infeld gravity (e.g. \cite{bana}).
Although, the Starobinsky model was originally proposed to explain inflation in the early universe, 
later, it has also been applied to describe neutron stars \cite{psaltis,eksi}.
Also, modified Starobinsky models, for example, with logarithmic and cubic corrections, have been used to obtain 
viable neutron star solutions \cite{capo1}. 
For the Starobinsky $f(R)$-model, the {\it modified} field equation is of the form
\begin{equation}
G_{\mu\nu} + \alpha \left[2 R G_{\mu\nu} + \frac{1}{2} R^2 g_{\mu\nu} - 2(\nabla_\mu \nabla_\nu - g_{\mu\nu}\Box)R \right] = 8\pi T_{\mu\nu} .
\label{modfld}
\end{equation}

\section{Solution Procedure}
\label{solnp}

Now that we have obtained the modified field equation, equation (\ref{modfld}), our next 
step would be to derive from it the corresponding {\it modified} 
Tolman-Oppenheimer-Volkoff (TOV) equations for this model and subsequently solve them to 
obtain the structure of the spherically symmetric white dwarf. Recall that the TOV equation in general relativity is obtained from 
equation (\ref{gr}).

\subsection{Perturbative method of solution and modified TOV equations}

Obtaining the modified TOV equations exactly from equation (\ref{modfld}) 
is quite complicated and laborious. Here we adopt a simpler and more intuitive way to deal 
with the problem, namely the perturbative method, 
which has been extensively applied in neutron stars \cite{psaltis,eksi,capo1,capo2,cheon,caponew}. 
So far we have not commented about the magnitude of $\alpha$ in equation (\ref{ar2}), except its 
astrophysical constraint. In the perturbative approach, $\alpha$ is considered to be a {\it small} 
parameter, such that $\alpha R \ll 1$. Thus the $\alpha R^2$ term in the Starobinsky model 
can be considered as a first order correction to general relativity, neglecting higher order corrections. 
Note that for $\alpha=0$, equation (\ref{modfld}) reduces to equation (\ref{gr}), giving back 
the zeroth-order results corresponding to general relativity.

Now, as a first step towards constructing the modified TOV equations in the perturbative approach, 
let us consider the spherically symmetric metric describing the interior of the star
\begin{eqnarray}
ds^2 &=& g_{\mu\nu}dx^\mu dx^\nu \nonumber \\
 &=& -e^{2\phi_\alpha} dt^2 + e^{2\lambda_\alpha} dr^2 + r^2(d\theta^2 + \sin^2 \theta d\phi^2) ,
\label{metric}
\end{eqnarray}
where $\phi_\alpha$ and $\lambda_\alpha$ are functions of the radial coordinate $r$. Note that 
perturbative constraint implies that $g_{\mu\nu}$ also has to be expanded in terms of $\alpha$ as 
$g_{\mu\nu} = g^{(0)}_{\mu\nu} + \alpha g^{(1)}_{\mu\nu} + {\cal O}(\alpha^2)$, where 
$g^{(0)}_{\mu\nu}$ is the metric in general relativity. 
We consider the matter source to be a perfect fluid described by 
\begin{equation}
T_{\mu\nu} = (\rho_\alpha + P_\alpha)u_\mu u_\nu + P_\alpha g_{\mu\nu} ,
\label{Tmunu}
\end{equation} 
where $\rho_\alpha$ is the density, $P_\alpha$ the pressure and $u_\mu$ the 4-velocity of the fluid. 
Again, for a perturbative solution we have $\rho_\alpha = \rho^{(0)} + \alpha \rho^{(1)}+ {\cal O}(\alpha^2)$ 
and $P_\alpha = P^{(0)} + \alpha P^{(1)}+ {\cal O}(\alpha^2) $. 
All the zeroth order quantities (e.g., $\phi^{(0)}$, $\lambda^{(0)}$, $\rho^{(0)}$ and $P^{(0)}$) 
correspond to the solution of Einstein's equations in general relativity. 
Taking all these into account and neglecting terms of ${\cal O}(\alpha^2)$ and higher, 
the temporal component ($\mu=\nu=t$) of equation (\ref{modfld}) yields
\begin{align}
-8\pi \rho_\alpha &= -r^{-2} + e^{-2\lambda_\alpha}(1-2r\lambda'_\alpha)r^{-2} + 
 \alpha \biggl[ 2 R^{(0)}(-r^{-2} + e^{-2\lambda^{(0)}}(1-2r\lambda^{(0)'})r^{-2}) \nonumber \\
& +\frac{1}{2} {R^{(0)}}^2 + 2 e^{-2\lambda^{(0)}}( R^{(0)'}r^{-1}(2-r\lambda^{(0)'}) +  R^{(0)''}) \biggr] ,
\label{rhoalp}
\end{align}
while the radial component ($\mu=\nu=r$) yields
\begin{align}
8\pi P_\alpha &= -r^{-2} + e^{-2\lambda_\alpha}(1+2r\phi'_\alpha)r^{-2}  + 
\alpha  \biggl[2R^{(0)}(-r^{-2} + e^{-2\lambda^{(0)}}(1+2r\phi^{(0)'})r^{-2}) \nonumber \\ &
+ \frac{1}{2} {R^{(0)}}^2 + 2 e^{-2\lambda^{(0)}}R^{(0)'}r^{-1}(2+r\phi^{(0)'}) \biggr] ,
\label{palp}
\end{align}
where prime $(')$ denotes single derivative with respect to $r$ and double prime $('')$ denotes 
double derivative with respect to $r$. 
Note that we are seeking perturbative solutions only up to
order $\alpha$ and, hence, for the terms already multiplied by $\alpha$, we invoke the zeroth 
order quantities $\lambda^{(0)}$, $\phi^{(0)}$ and $R^{(0)}$. 
The zeroth order Ricci scalar is defined as
\begin{equation}
R^{(0)} = 8\pi(\rho^{(0)} - 3P^{(0)}),
\label{ric0}
\end{equation}
which can be obtained by taking the trace of equation (\ref{gr}), when note that $R=R^{(0)}$ for equation (\ref{gr}). One can furthermore 
simplify equations (\ref{rhoalp}) and (\ref{palp}) by using the temporal and radial 
components of equation (\ref{gr}), given by, $-8\pi \rho^{(0)} = -r^{-2} + e^{-2\lambda^{(0)}}(1-2r\lambda^{(0)'})r^{-2}$ and 
$8\pi P^{(0)} = -r^{-2} + e^{-2\lambda^{(0)}}(1+2r\phi^{(0)'})r^{-2}$, respectively. 

Note that the exterior solution of the star is simply the vacuum solution of Einstein's 
equations which yields the Schwarzschild metric. Keeping that in mind we assume
\begin{equation}
e^{-2\lambda_\alpha} = 1 - \frac{2M_\alpha}{r},
\label{mpara}
\end{equation}
where $M_\alpha= M^{(0)} + \alpha M^{(1)} + {\cal O}(\alpha^2)$, is the mass of the star 
and $M^{(0)} = 4\pi \int \rho^{(0)} r^2 dr$, 
is the zeroth order mass (in general relativity), which corresponds to $e^{-2\lambda^{(0)}} = 1 - \frac{2M^{(0)}}{r}$. 
Using equation (\ref{mpara}) and its derivative in 
equation (\ref{rhoalp}), followed by some algebra, one obtains the mass equation
\begin{align}
\frac{dM_\alpha}{dr} &= 4\pi r^2\rho_\alpha -  \alpha \biggl[8\pi r^2 \rho^{(0)} R^{(0)} -\frac{c^2}{4G}{R^{(0)}}^2r^2 \nonumber \\ &
+ R^{(0)'}\left(4\pi r^3\rho^{(0)} + 3M^{(0)} - \frac{2c^2}{G}r \right) - \frac{c^2}{G}R^{(0)''}r^2\left(1-\frac{2GM^{(0)}}{c^2r}\right) \biggr] ,
\label{mass}
\end{align}
where we have plugged back $c$ and $G$ to make the equation dimensionful.

Next, from equation (\ref{palp}) we obtain the following dimensionful equation for the gravitational potential $\phi_\alpha (r)$
\begin{align}
\frac{d\phi_\alpha}{dr} &= \frac{G(\frac{4\pi r^3P_\alpha}{c^2} + M_\alpha)}{r^2(1-\frac{2GM_\alpha}{c^2r})} -
\frac{\alpha}{(1-\frac{2GM_\alpha}{c^2r})}\biggl[8\pi rR^{(0)} P^{(0)}\frac{G}{c^2} \nonumber \\ & 
- \frac{1}{4}c^2 r {R^{(0)}}^2 + R^{(0)'}\left(2c^2- \frac{3GM^{(0)}}{r} + 4\pi P^{(0)}r^2 \frac{G}{c^2} \right) \biggr] ,
\end{align}
which can be replaced in the equation of relativistic hydrostatic equilibrium
\begin{equation}
\frac{dP_\alpha}{dr} = - \left(\rho_\alpha + \frac{P_\alpha}{c^2} \right)\frac{d\phi_\alpha}{dr} ,
\label{equib}
\end{equation}
which is obtained from the conservation of the energy-momentum tensor, $g_{\nu r}\nabla_\mu T^{\mu \nu}=0$. 

Thus equations (\ref{mass}) and (\ref{equib}) together form the {\it modified} set of TOV equations, which 
reduce to the usual TOV equations in general relativity for $\alpha=0$. 
In this context, we mention that in the non-perturbative and exact approach, 
one can no longer invoke solutions of Einstein's equations as zeroth order terms, which is 
what makes that approach more difficult to handle numerically. Furthermore, if no perturbative 
constraints are imposed, then $\alpha$ can be arbitrarily large, which, however, might not 
respect astrophysically determined constraints obtained from perturbative calculations 
\cite{dam2,fre,naf}.

\subsection{Equation of state and boundary conditions}
\label{sec:eos}

In order to solve the modified TOV equations, one 
must also supply an equation of state (EoS)
relating the pressure and density within the star. 
In the current work, we use the EoS obtained 
by Chandrasekhar \cite{chandra35} for non-magnetized, non-rotating white dwarfs, which are constituted of electron 
degenerate matter. The pressure and density 
of such a system are respectively given by \cite{chandra35}
\begin{equation}
P_\alpha = \frac{\pi m_e^4c^5}{3h^3}[x(2x^2-3)\sqrt{x^2+1} + 3\sinh^{-1}x]
\label{Pchandra}
\end{equation}
and
\begin{equation}
\rho_\alpha = \frac{8\pi \mu_e m_H (m_e c)^3}{3h^3}x^3 ,
\label{rhochandra}
\end{equation}
where $x=p_F/(m_e c)$, $p_F$ is the Fermi momentum, $m_e$ the mass of electron, $h$ Planck's constant, 
$\mu_e$ the mean molecular weight per electron (we choose $\mu_e=2$ for our work) and $m_H$ the mass of hydrogen atom. 
Eliminating $x$ from equations (\ref{Pchandra}) and (\ref{rhochandra}) yields the EoS for the white dwarf.

Finally, the modified TOV equations, accompanied by the above EoS, can now be solved 
numerically, subjected to the boundary conditions $M_{\alpha}(r=0)=0$ and $\rho_\alpha(r=0)=\rho_c$, 
where $\rho_c$ is the central density of the white dwarf. Note that a particular $\rho_c$, supplied from the EoS, 
yields a particular mass $M_*$ and radius $R_*$ for a white dwarf. Hence, 
by varying $\rho_c$, one can construct the mass-radius relation for a given EoS. In the 
current work, we vary $\rho_c$ from $2\times 10^5$ g/$\rm cm^3$ to a maximum of $10^{11}$ g/$\rm cm^3$. 

\subsection{Relativistic Lane-Emden equations for modified gravity}

In Newtonian stellar structure theory, in order to capture a better physical insight, 
the hydrostatic equilibrium condition combined with 
Poisson's equation is recast into a dimensionless form for a polytropic fluid. This helps
in obtaining scaling relations for $M_*$ and $R_*$ with $\rho_c$ of the corresponding polytrope, 
known as the Lane-Emden formalism. 
With a similar aim, we apply the {\it relativistic} Lane-Emden formalism \cite{tooper} to the 
modified TOV equations and obtain analytical expressions for $M_*$ and $R_*$ (also see \cite{capomn}).
Here the hydrodynamic quantities are defined in terms of new dimensionless 
density and mass, $\theta$ and $\eta$, respectively. The zeroth order 
density in general relativity transforms as
\begin{equation}
\rho^{(0)} = \rho_c{\theta^{(0)}}^n, 
\end{equation}
where $n$ is the polytropic index. For the polytropic EoS
\begin{equation}
P^{(0)} = K {\rho^{(0)}}^\Gamma = K\rho_c^{1+(1/n)}{\theta^{(0)}}^{(n+1)} ,
\end{equation}
where $K$ is a constant and $\Gamma=1+(1/n)$. The radial coordinate transforms as
\begin{equation}
r = a\xi ,
\end{equation}
where $a$ has the dimension of length and is defined as
\begin{equation}
a= \left(\frac{(n+1)K\rho_c^{(1-n)/n}}{4\pi G}\right)^{1/2} .
\end{equation}
The zeroth order mass in general relativity transforms as
\begin{equation}
M^{(0)} = 4\pi \rho_c a^3 \eta^{(0)} .
\end{equation}
Similarly, all the hydrodynamic quantities in our chosen modified gravity theory also transform as follows:
\begin{equation}
\rho_\alpha = \rho_c \theta_\alpha^n, 
\end{equation}
\begin{equation}
P_\alpha = K \rho_\alpha^\Gamma = K\rho_c^{1+(1/n)}\theta_\alpha^{(n+1)} ,
\end{equation}
and
\begin{equation}
M_\alpha = 4\pi \rho_c a^3 \eta_\alpha .
\end{equation}
Thus, in terms of the new dimensionless variables, the two modified TOV equations 
can be cast into the modified relativistic Lane-Emden form. Equation (\ref{mass}) becomes
\begin{align}
\frac{d\eta_\alpha}{d\xi} = I_m &= \xi^2 \theta_\alpha^n - \alpha R^{(0)} \biggl[2\xi^2{\theta^{(0)}}^n - 
\frac{\xi^2{\theta^{(0)}}^n}{2}(1-3\sigma\theta^{(0)}) \nonumber \\ & 
-\frac{1}{R^{(0)}}\frac{dR^{(0)}}{d\xi} \left(\frac{2\xi}{(n+1)\sigma} -3\eta^{(0)}- \xi^3{\theta^{(0)}}^n \right) 
-\frac{\xi^2}{(n+1)\sigma}\frac{1}{R^{(0)}}\frac{d^2 R^{(0)}}{d\xi^2}\left(1 - \frac{2\eta^{(0)}(n+1)\sigma}{\xi}  \right) \biggr] ,
\label{lane_mass}
\end{align}
while equation (\ref{equib}) becomes
\begin{align}
\frac{d\theta_\alpha}{d\xi} &= -\frac{(1+\sigma\theta_\alpha)(\eta_\alpha+\xi^3\sigma\theta_\alpha^{(n+1)})}{\xi^2(1-\frac{2\eta_\alpha(n+1)\sigma}{\xi})} 
+ \frac{\alpha R^{(0)}(1+\sigma\theta_\alpha)}{(1-\frac{2\eta_\alpha(n+1)\sigma}{\xi})}\biggl[ 2\xi\sigma{\theta^{(0)}}^{(n+1)} 
-\frac{\xi{\theta^{(0)}}^n(1-3\sigma\theta^{(0)})}{2} \nonumber \\ & + \frac{1}{R^{(0)}}\frac{dR^{(0)}}{d\xi}\left( \frac{2}{\sigma(n+1)} - \frac{3\eta^{(0)}}{\xi} + \xi^2\sigma{\theta^{(0)}}^{(n+1)}   \right)       \biggr] ,
\end{align}
where $\sigma=P_c/(\rho_c c^2)$, $P_c$ being the central pressure of the star. The boundary conditions required to solve these equations 
are $\theta_\alpha(\xi=0)=1$ and $\eta_\alpha(\xi=0)=0$. Note that $\alpha R^{(0)}$ is a dimensionless quantity 
and $R^{(0)}= \frac{8\pi G}{c^2}\rho_c {\theta^{(0)}}^n (1-3\sigma \theta^{(0)})$. 
Furthermore, note that for $\alpha=0$, the above equations reduce to the relativistic Lane-Emden equations corresponding to the TOV equations in general relativity, 
whereas for $\alpha=\sigma=0$ we obtain the Lane-Emden equations for a Newtonian system.

The radius $R_*$ of the star is given by
\begin{equation}
R_* = a \xi_1 = \left(\frac{(n+1)K}{4\pi G}\right)^{1/2} \rho_c^{(1-n)/2n} \xi_1 ,
\end{equation}
where $\xi_1$ corresponds to the first zero of the function $\theta_\alpha (\xi)$.
The mass  $M_*$ of the star is given by
\begin{equation}
M_* = 4\pi \left(\frac{(n+1)K}{4\pi G}\right)^{3/2} \rho_c^{(3-n)/2n} \eta_\alpha(\xi_1) ,
\end{equation}
where $\eta_\alpha (\xi_1) = \int_0^{\xi_1} I_m d\xi$ (see equation \ref{lane_mass}).

Now, for high density ($\rho_c \gtrsim 5\times 10^9$ g/$\rm cm^3$), relativistic white dwarfs, the EoS associated with equations (\ref{Pchandra}) and (\ref{rhochandra}) 
can be simply described by a $n=3$ polytropic EoS with $K=(1/8)(3/\pi)^{1/3} hc/(\mu_e m_H)^{4/3}$. 
The mass and radius for such white dwarfs hence become
\begin{equation}
M_* = 4\pi \left(\frac{K}{\pi G}\right)^{3/2} \eta_\alpha(\xi_1)
\end{equation}
and
\begin{equation}
R_* = \left(\frac{K}{\pi G}\right)^{1/2} \rho_c^{-1/3}\xi_1 .
\end{equation}

Note that in the corresponding Newtonian case, $M_*$ is completely independent of $\rho_c$, giving rise to the limiting mass. 
However, in both general relativity and modified gravity, $M_*$ implicitly depends on $\rho_c$ through the parameter $\sigma$, which determines $\eta_\alpha(\xi_1)$.

\section{Results}
\label{res}

\begin{figure}[]
\begin{center}
\includegraphics[angle=0,width=18cm]{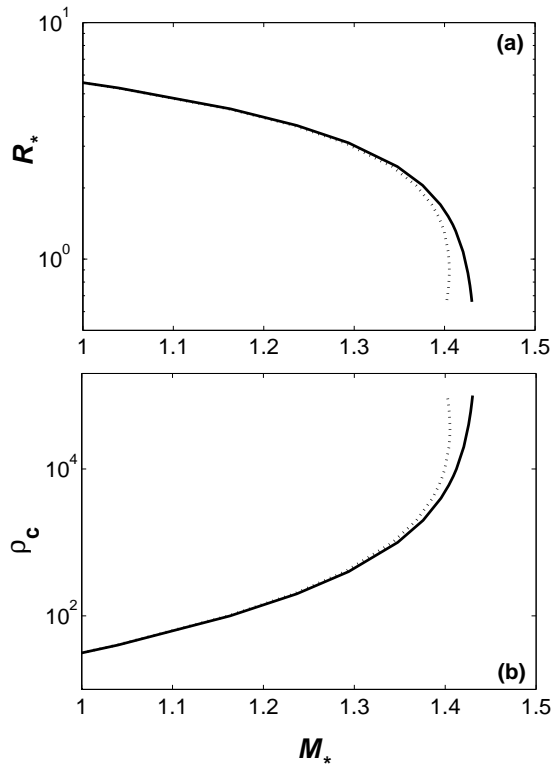}
\caption{Comparison of solutions in the Newtonian case (solid lines) and the general 
relativistic or $\alpha=0$ case (dotted lines). 
(a) Mass-radius relations. (b) Variation of $\rho_c$ with $M_*$. 
$\rho_c$, $M_*$ and $R_*$ are in units of $10^6$ g/$\rm cm^3$, $M_\odot$ and 1000 km, respectively.
  }
\label{comp}
\end{center}
\end{figure}

\begin{figure}[]
\begin{center}
\includegraphics[angle=0,width=18cm]{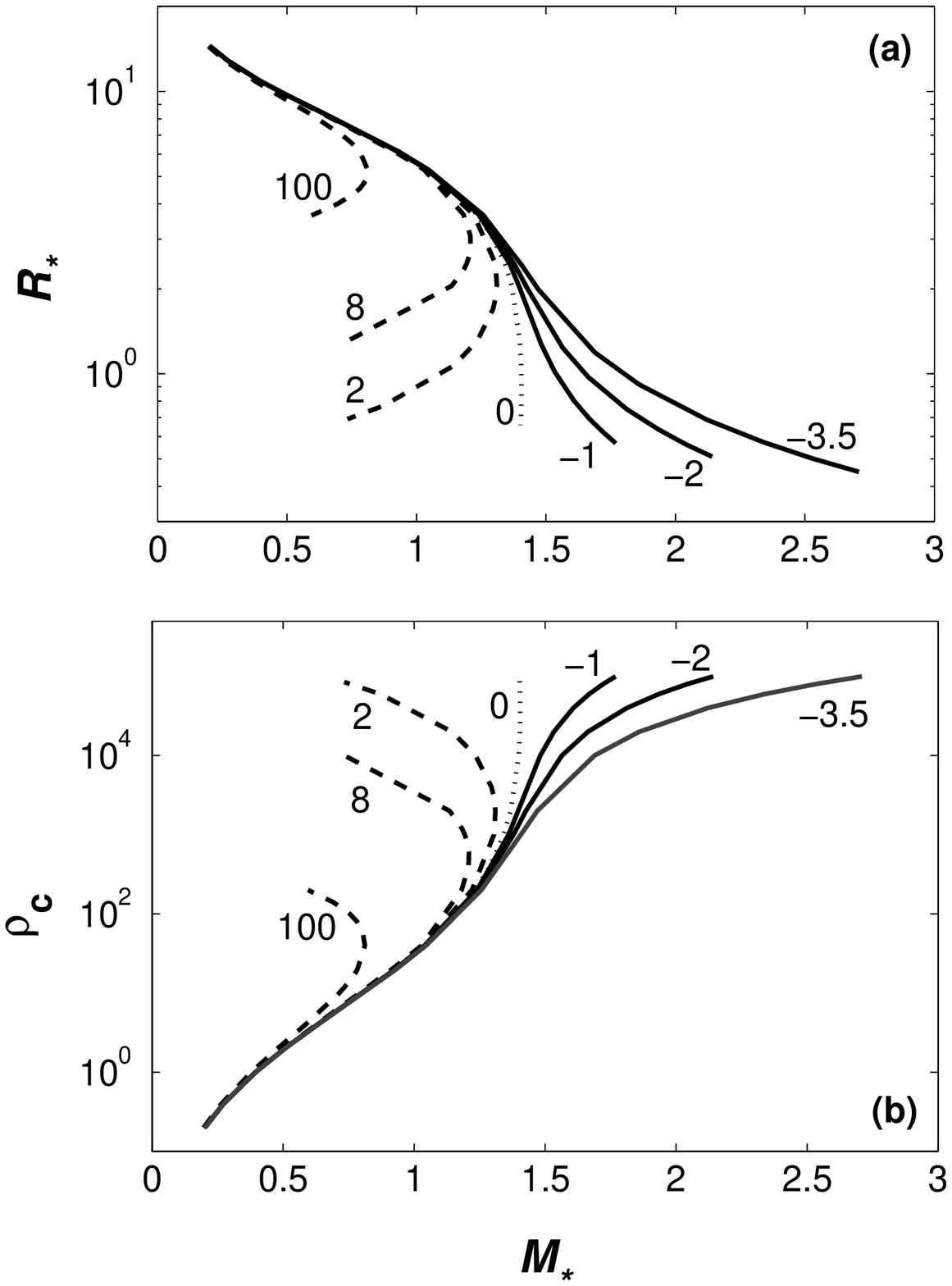}
\caption{Unification diagram for SNeIa.
(a) Mass-radius relations. (b) Variation of $\rho_c$ with $M_*$.
The numbers adjacent to the various lines denote $\alpha/(10^{13}~\rm cm^2)$. 
$\rho_c$, $M_*$ and $R_*$ are in units of $10^6$ g/$\rm cm^3$, $M_\odot$ and 1000 km, respectively.
  }
\label{uni}
\end{center}
\end{figure}

Finally, we move on to describe the results obtained from our calculations, which are illustrated in Figure 
\ref{comp}, for $\alpha=0$ (confirming known results) and in Figure \ref{uni}, for different values of $\alpha$.
The choice of different $\alpha$-s and corresponding different results imply the variation of 
$\alpha$ with space (and/or space-time). A more physical model
should automatically account for the variation of $\alpha$ or equivalent parameter(s) with the 
density of white dwarfs. The present work just argues for different results at different $\alpha$-s
corresponding to the different regimes of density. In future, we should repeat the present work 
using a viable $f(R)$ model (e.g. \cite{cham1,cham2}) 
consistent with solar system constraints, which reveals the density dependent modification to the 
gravity effects, by exploiting, e.g., the so-called chameleon effect.

\subsection{Case with $\alpha=0$}

In Figure \ref{comp}, we recall the well known results where we compare 
the Newtonian solutions with those in general relativity, i.e., the $\alpha=0$ case.
Figures \ref{comp}(a) and (b) confirm that in the Newtonian case, with the increase of $\rho_c$, 
$R_*$ decreases and $M_*$ increases, until it saturates to a maximum mass $M_{\rm max} \sim 1.44M_\odot$, 
which is the famous Chandrasekhar limit. We furthermore confirm that in the general relativistic case, with the increase of $\rho_c$, 
$R_*$ decreases but $M_*$ increases until it reaches $M_{\rm max}=1.405M_\odot$ 
at $\rho_c=3.5\times 10^{10}$ g/$\rm cm^3$. 
A further increase in $\rho_c$ results in a slight decrease in $M_*$, indicating the onset of an unstable branch, 
which is absent in the Newtonian case \cite{chandra35}. 
For low density white dwarfs having $\rho_c < 10^8$ g/$\rm cm^3$, the Newtonian and general relativistic $M_*-\rho_c$ 
curves are identical. However, for $\rho_c \gtrsim 10^8$ g/$\rm cm^3$, general relativistic 
effects become important, resulting in a slightly smaller $M_*$ compared to the 
Newtonian case and eventually leading to a smaller $M_{\rm max}$. Once this $M_{\rm max}$ is approached,
by further gaining mass, white dwarfs contract, causing an increase in the core temperature 
and, finally, leading to runaway thermonuclear reactions, which result in SNeIa.

\subsection{Cases with $\alpha<0$}

Let us now consider the $\alpha<0$ cases, with $\alpha$ much less than the strictest astrophysical upper bound.
Figure \ref{uni}(b) shows that  
for $\rho_c < 10^8$ g/$\rm cm^3$, all the three $M_*-\rho_c$ curves 
are indistinguishable from the $\alpha=0$ case (note that the $\alpha=0$ curves in Figure \ref{uni} are 
identical to the dotted lines in Figure \ref{comp}). As $\rho_c$ increases beyond $10^8$ g/$\rm cm^3$, the 
curves deviate more and more due to modified gravity effects. This feature beautifully 
establishes the necessary constraint that a modified gravity theory should replicate general relativistic 
results in the appropriate regime, which for white dwarfs is the low density regime. 
This furthermore, very importantly,
reveals that modified gravity has a tremendous impact on white dwarfs which so far was completely overlooked, 
whereas general relativistic effect itself is non-negligible.
Note that the values of $M_{\rm max}$ for all the three cases correspond to $\rho_c=10^{11}$ g/$\rm cm^3$,
an upper limit chosen to avoid possible neutronization. 
Interestingly, all values of $M_{\rm max}$ are highly super-Chandrasekhar, ranging from 
$1.8-2.7M_\odot$. The corresponding values of $\rho_c$ are large enough to initiate thermonuclear reactions, e.g.
they are larger than $\rho_c$ corresponding to $M_{\rm max}$ of $\alpha=0$ case, whereas the core temperatures of 
the respective limiting mass white dwarfs are expected to be similar.This could
explain the entire range of the observed super-SNeIa mentioned above \cite{howel,scalzo,taub,yam,hicken,silverman}. 
While the general relativistic effect is very small, modified gravity effect could, according to the perturbative 
$f(R)$-model, lead to $\sim 100\%$ increase in the limiting mass of white dwarfs. 
Similarly, in case of neutron stars, small values of a corresponding $\alpha$-equivalent parameter 
were shown to reveal large deviations in their mass \cite{dam}. The modified gravity effect particularly is
pronounced at the high density regime, even if $\alpha$ is very small. 
We also find that unlike the $\alpha=0$ case, the $M_*-R_*$ relations for white dwarfs having $\alpha<0$
consist of only a stable branch, i.e., as $\rho_c$ increases, $M_*$ 
always increases as seen in Figure \ref{uni}(b). 
The results of the Lane-Emden solutions 
for several $\alpha<0$ cases are listed in Table \ref{table1} for $\rho_c=10^{11}$ g/$\rm cm^3$, which yields $M_*=M_{\rm max}$.
We observe that $M_{\rm max}$ (and the corresponding $R_*$) for the three $\alpha<0$ cases shown in Figure \ref{uni} (obtained 
directly from solving the modified TOV equations) agree with the values in Table \ref{table1}.

\begin{table}[]
\caption{Maximum mass from relativistic Lane-Emden solutions for modified gravity with $\alpha<0$, $n=3$ and $\sigma=2.527\times 10^{-3}$, where $\alpha_{13}=\alpha/(10^{13}~ \rm cm^2)$.}
\begin{center}
\small
\begin{tabular}{|c|c|c|c|c|c|c|c|}
\hline 
$\alpha_{13}$  & $\xi_1$ & $\eta_\alpha(\xi_1)$ & $M_{\max}$ ($M_\odot$)& $R_*$ (1000 km) & 
$|\alpha R^{(0)}|_{\rm max}$ & $|1-g^{(0)}_{tt}/g_{tt}|_{\rm max}$ & $|1-g^{(0)}_{rr}/g_{rr}|_{\rm max}$ \\ 
\hline
-1 & 5.7832 & 2.49083 & 1.772 & 0.6027 & 0.00184 & 0.0016 & 0.0052 \\ 
-2 & 5.1032 & 3.00800 & 2.139 & 0.5318 & 0.00369 & 0.0031 & 0.0108 \\
-2.5 & 4.8507 & 3.26938 & 2.325 & 0.5055 & 0.00462 & 0.0038 & 0.0138 \\ 
-3 & 4.6375 & 3.53262 & 2.513 & 0.4833 & 0.00554 & 0.0045 & 0.0168 \\
-3.5 & 4.4547 & 3.79753 & 2.701 & 0.4643 & 0.00646 & 0.0052 & 0.0199 \\ \hline
\end{tabular}
\end{center}
\label{table1}
\end{table}

The last three columns of Table \ref{table1} list three extra parameters that give a measure for ensuring 
the perturbative validity of the solutions for a chosen $\alpha$, which we hereby define. Recall that 
we solve the modified TOV equations only up to ${\cal O} (\alpha)$ and since the product 
$\alpha R$ is first order in $\alpha$, we replace $R$ in it by $R^{(0)}$, given by equation (\ref{ric0}). 
The maximum value of $|\alpha R^{(0)}|_{\rm max}$
occurs at the center of the white dwarf, and for the perturbative validity of the entire solution,  $|\alpha R^{(0)}|_{\rm max} \ll 1$ should hold 
true. Next we consider the ratios $g_{tt}^{(0)}/g_{tt}$ and $g_{rr}^{(0)}/g_{rr}$,
which should be close to 1 for the validity of perturbative approach \cite{orelana}. We consider the maximum 
deviation of these quantities from 1, such that $|1-g_{tt}^{(0)}/g_{tt}|_{\rm max} \ll 1$ and 
$|1-g_{rr}^{(0)}/g_{rr}|_{\rm max} \ll 1$ should both hold true.
Table \ref{table1} shows that all the three error estimates are at least $2-3$ 
orders of magnitude smaller than 1 and chosen $\alpha$-s are perfectly in accordance with the observational 
constraints \cite{dam2,fre,naf}. 
Note, furthermore, that there is no universal quantity which gives an absolute measure of the allowed deviation 
from general relativity in the perturbative approach and, hence, we discuss above at least three 
such possible quantities. An additional estimate of the error may be obtained from a quantity defined in 
\cite{psaltis}, which we denote here as, $\delta_P = (dP_\alpha/dr)/(dP^{(0)}/dr) -1$. For 
perturbative validity of the solution $|\delta_P| \lesssim 1$, a condition satisfied for 
the cases listed in Table \ref{table1}.

\subsection{Cases with $\alpha>0$}

Coming to the $\alpha>0$ cases, Figure \ref{uni}(b) shows 
that all the three $M_*-\rho_c$ curves overlap with the $\alpha=0$ curve in the low density region. 
However, with the increase 
in the magnitude of $\alpha$, the region of overlap recedes to a lower $\rho_c$. Modified gravity 
effects set in at $\rho_c \gtrsim 10^8,~4\times 10^7$ and $2\times 10^6$ g/$\rm cm^3$, 
for $\alpha = 2\times 10^{13}~ {\rm cm^2}$, $8\times 10^{13}~ {\rm cm^2}$ and 
$10^{15}~{\rm cm^2}$ respectively. For a given $\alpha$, with the increase of $\rho_c$, $M_*$ 
first increases, reaches a maximum ($M_{\rm max}$) and then decreases, like the $\alpha=0$ case. With the 
increase of $\alpha$, $M_{\rm max}$ decreases and, interestingly,
for $\alpha=10^{15}~{\rm cm}^2$, it is highly sub-Chandrasekhar ($0.81M_\odot$).
In fact, $M_{\rm max}$ for all the chosen $\alpha>0$ is sub-Chandrasekhar, ranging $1.31-0.81M_\odot$. 
This is a remarkable finding since it establishes that even if the values of $\rho_c$ for 
these sub-Chandrasekhar maximum/limiting mass white dwarfs are lower than the conventional value at which 
SNeIa are usually triggered, an attempt 
to increase the mass beyond $M_{\rm max}$, for a given $\alpha$, will lead to a gravitational instability. 
This presumably will be followed by a 
runaway thermonuclear reaction, provided the core temperature increases sufficiently due to collapse. 
One might wonder if such low density sub-Chandrasekhar limiting mass white dwarfs can attain 
conditions suitable to initiate a detonation, which would give way to a SNIa. Interestingly, the 
occurrence of such a detonation has already been demonstrated in white dwarfs having densities as low as 
$\sim 10^6$ g/$\rm cm^3$, provided certain background conditions are satisfied \cite{runaway}.
Thus, once the maximum mass is approached, a SNIa is expected to trigger just 
like in the $\alpha=0$ case. The explosions of these sub-Chandrasekhar white dwarfs 
could explain the sub-SNeIa \cite{1991bg,mazzali97,turato97,modjaz,garnavich,taub2008}, 
like SN 1991bg mentioned above, because a small progenitor 
mass will consequently yield a small Ni mass leading to an under-luminous event. 
Note that, as evident from Figure \ref{uni}(b), the $M_*-\rho_c$ curves for the $\alpha>0$ cases 
terminate at different $\rho_c$s unlike the $\alpha<0$ cases. 
This is because, when $\rho_c$ exceeds a certain value for a given positive $\alpha$, the numerical/mathematical
solutions reveal a region of 
negative mass within the white dwarf with an overall positive $M_*$. With a further increase in 
$\rho_c$, the entire $M_*$ becomes negative. These are unphysical scenarios and, hence, 
in Figure \ref{uni}, we present the $\alpha>0$ solutions only up to that $\rho_c$ for 
which the mass is positive throughout the white dwarf. 

We now check the validity of the perturbative approach for the $\alpha>0$ cases, 
corresponding to the respective $M_{\rm max}$. 
For $\alpha=2\times 10^{13}~{\rm cm}^2$, $|\alpha R^{(0)}|_{\rm max}=7.4\times10^{-5}$, 
$|1-g_{tt}^{(0)}/g_{tt}|_{\rm max} =  6.8\times 10^{-5}$ and 
$|1-g_{rr}^{(0)}/g_{rr}|_{\rm max} = 2\times 10^{-4}$; 
for $\alpha=8\times 10^{13}~{\rm cm}^2$, $|\alpha R^{(0)}|_{\rm max}=7.4\times10^{-5}$, 
$|1-g_{tt}^{(0)}/g_{tt}|_{\rm max} =  6.8\times 10^{-5}$ and 
$|1-g_{rr}^{(0)}/g_{rr}|_{\rm max} = 2.0\times 10^{-4}$; 
and for $\alpha= 10^{15}~{\rm cm}^2$, $|\alpha R^{(0)}|_{\rm max}=7.4\times10^{-5}$, 
$|1-g_{tt}^{(0)}/g_{tt}|_{\rm max} =  6.9\times 10^{-5}$ and 
$|1-g_{rr}^{(0)}/g_{rr}|_{\rm max} = 2\times 10^{-4}$. 
This ensures that the solutions are within 
the perturbative regime and are perfectly in accordance with the observational constraints \cite{dam2,fre,naf}. 
Also, $|\delta_P|<1$ for all the above $\alpha>0$ cases.

In this context, we mention for comparison the results of a recent work on the application of 
Starobinsky gravity in neutron stars \cite{bul}, which adopts a fully non-perturbative method. It 
reports that smaller positive values of $\alpha$ (perturbative limit) lead to a decrease in the mass of 
neutron stars, while larger positive values of $\alpha$ (non-perturbative regime) lead to an increase in 
mass, with respect to that in the general relativistic case. 
Our preliminary calculation, which is beyond the scope of the current paper, shows
a similar trend in white dwarfs. In future, we plan to report the non-perturbative results for 
white dwarfs based on established numerical techniques \cite{babi1,babi2}.

\section{Conclusions}
\label{conc}

Based on a simple $f(R)$-model, we show, for the first time in the literature to the best 
of our knowledge, that modified gravity effects are significant
in high density white dwarfs. Consideration of such effects in white dwarfs appears to be 
indispensable, since it appears to be remarkably explaining and  
unifying a wide range of observations for which general relativity may be insufficient.
Importantly, we are also able to show that the $f(R)$-model chosen in our work successfully 
reproduces the low density white dwarfs and their basic properties, which are already explained in the 
paradigm of general relativity (and Newtonian framework).

We note here that the perturbative method is adequate for 
the present study, as then we have a handle on $\alpha$ characterizing our model, which cannot be 
arbitrarily large,
allowing it to be constrained directly by astrophysical observations. 
In our work, for the super-Chandrasekhar 
limiting mass white dwarfs, $\alpha$ ranges from $-10^{13}$ to $-3.5\times 10^{13}$ ${\rm cm^2}$, while for the sub-Chandrasekhar 
limiting mass white dwarfs, the range is $2\times 10^{13}$ to $10^{15}$ ${\rm cm^2}$. Hence, the range of $\alpha$ chosen in our work 
is well within the astrophysical bound set by the Gravity Probe B experiment, 
namely $|\alpha|\lesssim 5\times 10^{15}$ ${\rm cm^2}$ \cite{naf}.

Furthermore, even though $\alpha$ is assumed to be constant within 
individual white dwarfs here, there is indeed an implicit 
dependence of $\alpha$ on the central density, particularly of the limiting mass white dwarfs presumably leading to 
SNeIa, as is evident from Figure \ref{uni}(b). This indicates the existence of a chameleon-like effect 
in observed SNIa progenitors. A more sophisticated calculation, which invokes an (effective) 
$\alpha$ that varies explicitly with density, is likely to yield results similar to those we have 
already obtained in this work.

Depending on the magnitude and sign of $\alpha$, we are not only able to obtain both 
highly super-Chandrasekhar (for $\alpha<0$) and highly sub-Chandrasekhar (for $\alpha>0$) 
limiting mass white dwarfs, but we can also establish them as progenitors of the peculiar, super-SNeIa and 
sub-SNeIa, respectively. Thus, an effectively single underlying theory, 
inspired by the need to modify Einstein's theory of general relativity, 
appears to be able to unify the two apparently disjoint sub-classes of SNeIa, and, hence, serves as a missing link,
which have so far hugely puzzled astronomers. The significance of the current work lies in the 
fact that it not only questions the uniqueness of the Chandrasekhar mass-limit for white dwarfs, 
but it also argues for the need of a modified theory of gravity to explain astrophysical 
observations.

\acknowledgments

B.M. acknowledges partial support through research Grant No. ISRO/RES/2/367/10-11. 
U.D. thanks CSIR, India for financial support. The authors would like to thank 
K. Y. Ek\c{s}i for useful discussion. Thanks are also due to the anonymous referee
for the suggestions to improve the presentation of the paper.

\end{document}